\documentclass[11pt]{article}
\usepackage{amssymb,latexsym,amsmath}
\usepackage[dvips]{graphicx}
\usepackage{cite}
\newtheorem{theo}{Theorem}

\newtheorem{prop}[theo]{Proposition}
\def\ii{\mathrm{i}} % imaginary unit i
\def\ee{\mathrm{e}} % exponential unit e
\def\su{\mathfrak{su}}

\def\gl{\mathfrak{gl}}
\def\ssl{\mathfrak{sl}}
\def\osp{\mathfrak{osp}}

\def\qdots{\mathinner{\mkern1mu\raise1pt\vbox{\kern7pt\hbox{.}}\mkern2mu
 \raise4pt\hbox{.}\mkern2mu\raise7pt\hbox{.}\mkern1mu}}
\newcommand{\myatop}[2]{\genfrac{}{}{0pt}{}{#1}{#2}}
\def\mybox{\hfill$\Box$}
\begin{document}
\begin{center}
{\Large \bf
Discrete series representations for $\ssl(2|1)$, Meixner polynomials and oscillator models}\\[5mm]
{\bf E.I.\ Jafarov\footnote{Permanent address: 
Institute of Physics, Azerbaijan National Academy of Sciences, Javid av.\ 33, AZ-1143 Baku, Azerbaijan}
and J.\ Van der Jeugt} \\[1mm]
Department of Applied Mathematics and Computer Science,
Ghent University,\\
Krijgslaan 281-S9, B-9000 Gent, Belgium\\[1mm]
E-mail: ejafarov@physics.ab.az, Joris.VanderJeugt@UGent.be
\end{center}

\vskip 10mm
\noindent
Short title: $\ssl(2|1)$ Meixner oscillator


%\addtolength{\baselineskip}{2mm}
%\addtolength{\abovedisplayskip}{1mm}
%\addtolength{\belowdisplayskip}{1mm}
%\addtolength{\parskip}{1mm}

\begin{abstract}
We explore a model for the one-dimensional quantum oscillator based upon the Lie superalgebra $\ssl(2|1)$.
For this purpose, a class of discrete series representations of $\ssl(2|1)$ is constructed, each representation
characterized by a real number $\beta >0$. 
In this model, the position and momentum operators of the oscillator are odd elements of $\ssl(2|1)$ and their expressions
involve an arbitrary parameter $\gamma$.
In each representation, the spectrum of the Hamiltonian is the same as that of the canonical oscillator.
The spectrum of the position operator can be continuous or infinite discrete, depending on the value of $\gamma$.
We determine the position wavefunctions both in the continuous and discrete case, and discuss their properties.
In the discrete case, these wavefunctions are given in terms of Meixner polynomials.
{}From the embedding $\osp(1|2)\subset\ssl(2|1)$, it can be seen why the case $\gamma=1$ corresponds
to the paraboson oscillator. Consequently, taking the values $(\beta,\gamma)=(1/2,1)$ in the
$\ssl(2|1)$ model yields the canonical oscillator.
\end{abstract}

\section{Introduction}

The quantum harmonic oscillator is one of the main examples in physics, both because of its use in physical models
and approximations and because of its mathematical beauty.
Besides the canonical oscillator there are many non-canonical versions that received attention in the literature.
In particular, there are many algebraic constructions to model a quantum oscillator by extending (or deforming) the
common oscillator Lie algebra.
The difficulty for such models is often to determine the spectra of observables and an explicit form of 
their eigenfunctions. Only for some models, one can develop such a complete theory.
One of these models is the $q$-oscillator, a $q$-deformation of the standard quantum 
oscillator~\cite{Macfarlane1989,Biedenharn1989,Sun1989}, discussed in this
context in~\cite{Arik1999} and~\cite{Klimyk2005}.

During the last decade, new oscillator models were developed such that the same dynamics as in the 
classical or quantum case is satisfied, and in such a way that the 
operators corresponding to position, momentum and Hamiltonian are elements of some algebra different
from the traditional Heisenberg (or oscillator) Lie algebra.
In the one-dimensional case, there are three (essentially self-adjoint) operators involved: 
the position operator $\hat q$, its corresponding momentum operator $\hat p$ and
the Hamiltonian $\hat H$ which is the generator of time evolution. 
The main requirement is that these operators should satisfy the Hamilton-Lie equations 
(or the compatibility of Hamilton's equations with the Heisenberg equations):
\begin{equation}
[\hat H, \hat q] = -\ii \hat p, \qquad [\hat H,\hat p] = \ii \hat q,
\label{Hqp}
\end{equation}
in units with mass and frequency both equal to~1, and $\hbar=1$.
Contrary to the canonical case, the commutator $[\hat q, \hat p]=\ii$ is not required. 
Apart from~\eqref{Hqp} and the self-adjointness, it is then common to require the following conditions~\cite{Atak2001}:
\begin{itemize}
\item all operators $\hat q$, $\hat p$, $\hat H$ belong to some Lie algebra or Lie superalgebra $\cal A$;
\item the spectrum of $\hat H$ in (unitary) representations of $\cal A$ is equidistant.
\end{itemize}

The model that has received most attention occurs for ${\cal A}= \su(2)$ (or its enveloping algebra)~\cite{Atak2001,Atak2001b,Atak2005}.
For this case, the appropriate representations are the common $\su(2)$ representations 
labeled by an integer or half-integer $j$. 
Since these representations are finite-dimensional, one is dealing with ``finite oscillator models'', of
potential use in optical image processing~\cite{Atak2005}.
Recently, this model has been extended by introducing an additional parameter in the algebra,
thus leading to finite paraboson oscillator models~\cite{JSV2011,JSV2011b}.
In~\cite{JV2012}, the Lie algebra $\su(2)$ was extended to the Lie superalgebra $\ssl(2|1)$,
yielding another interesting finite oscillator model with appealing position spectra and
remarkable discrete wavefunctions.

All the examples just mentioned concern finite oscillator models, with a finite (equidistant) spectrum for the 
Hamiltonian and with a discrete spectrum for the position and momentum operator, due to the
fact that the relevant (unitary) representations of the model algebra ${\cal A}$ are finite-dimensional.
But there are also some models of quantum oscillators with continuous spectra of position and momentum operators 
available besides the canonical model. 
One model is based on the positive discrete series representations of $\su(1,1)$~\cite{Klimyk2006}.
In such a representation the spectrum of the position operator is ${\mathbb{R}}$. 
The position wavefunctions are given by normalized Meixner-Pollaczek polynomials. 
Klimyk discussed many fascinating properties of these $\su(1,1)$ oscillators~\cite{Klimyk2006}.
A $q$-deformation of Klimyk's $\su(1,1)$ model was investigated in~\cite{Atak2006}. 
The position and momentum operators have spectra covered by a finite interval of the real line, which depends on
the value of $q$, and the wavefunctions are given in terms of $q$-Meixner-Pollaczek polynomials.
Another extension of the Lie algebra $\su(1,1)$ by means of a parity or reflection operator $R$ was
studied in~\cite{JSV2012}, together with a class of discrete series representations.
In the corresponding model, the Hamiltonian has a discrete but infinite equidistant spectrum, 
and the position operator has spectrum ${\mathbb{R}}$. Again, the model is sufficiently simple to construct
the position wavefunctions explicitly, and these are given in terms of continuous 
dual Hahn polynomials~\cite{Koekoek, Ismail}.

In the current paper we return to the Lie superalgebra ${\cal A}=\ssl(2|1)$, but we shall consider
a new class of infinite-dimensional discrete series representations labeled by a positive real number $\beta$. 
In the model using these representations, the choice for the Hamiltonian $\hat H$ is unambiguous and
(as required) it has a discrete but infinite equidistant spectrum.
The choice for the position operator $\hat q$ is not completely fixed: an arbitrary real parameter $\gamma$
remains in the expression for $\hat q$ in terms of the $\ssl(2|1)$ generators.
The momentum operator $\hat p$ follows from the first equation of~\eqref{Hqp}.
Our main work is then devoted to determining the spectrum of $\hat q$ and of its formal eigenvectors
from which the position wavefunctions $\Phi_n^{(\beta,\gamma)}(x)$ follow.
Our analysis shows that $\hat q$ has an infinite discrete spectrum when $|\gamma|\ne 1$, but a continuous
spectrum when $|\gamma|=1$. 
It is quite remarkable to have both of these situations occurring in the same oscillator model.
In the case $\gamma=1$, the position wavefunctions coincide with those of the paraboson oscillator
(i.e.\ they are given in terms of generalized Laguerre polynomials),
and in particular for $\beta=1/2$ the model coincides with the canonical oscillator.
When $|\gamma|\ne 1$, the position wavefunctions are given in terms of Meixner polynomials,
a class of orthogonal polynomials with a discrete orthogonality relation.

The structure of the paper is as follows.
In the next section, we give the Lie superalgebra $\ssl(2|1)$ and show that it possesses a class of 
discrete series representations. 
In section~3, the $\ssl(2|1)$ oscillator model is presented. 
In particular, using the technique of unbounded Jacobi matrices, the spectrum of the 
position and momentum operator is determined. 
Section~4 is devoted to studying the shape of the position wavefunctions.
We examine some plots of wavefunctions, and investigate how the discrete wavefunctions for $\gamma\ne 1$ 
(given in terms of Meixner polynomials) tend to
the continuous wavefunctions for $\gamma=1$ (given in terms of Laguerre polynomials),
both in plots and as a limit computation.
In section~5, we determine the corresponding $\ssl(2|1)$ Fourier transform, which is defined as the
kernel relating position eigenvectors to momentum eigenvectors.
Due to the fact that bilinear generating functions are known for the Meixner polynomials appearing here,
the $\ssl(2|1)$ Fourier transform can be given in explicit form.
In section~6 we briefly recall the paraboson oscillator model and its relation to the Lie algebra $\osp(1|2)$.
The embedding $\osp(1|2) \subset \ssl(1|2)$ explains algebraically why $\gamma=1$ is a special case in our model and
why it corresponds to the paraboson oscillator.
Finally, some remarks and a further discussion of some interesting quantities in the model is presented 
in section~7.

\section{The Lie superalgebra $\ssl(2|1)$ and a class of discrete series representations}

The Lie superalgebra $\ssl(2|1)$ with even part $\ssl(2)\oplus \gl(1)$ is well known, 
and has been constructed in a previous paper~\cite{JV2012}
where a class of finite-dimensional representations were used.
Let us just recall the basis here, following the choice of~\cite[p.~261]{Frappat}, consisting of 
four odd (or `fermionic') basis elements $F^+, F^-, G^+, G^-$ and four even
(or `bosonic') basis elements $H, E^+, E^-, Z$, given by
\begin{align}
& F^+=e_{32},\ G^+=e_{13},\ F^-=e_{31},\ G^-=e_{23}, \label{odd-el}\\
& H=\frac12(e_{11}-e_{22}),\ E^+=e_{12},\ E^-=e_{21},\ Z=\frac12(e_{11}+e_{22})+e_{33} \label{even-el}
\end{align}
in terms (graded) $3\times 3$ Weyl matrices $e_{ij}$.
The basis for the $\ssl(2)$ subalgebra is $\{H,E^+,E^-\}$ and 
the $\gl(1)\cong U(1)$ subalgebra is spanned by $Z$.
The basic Lie superalgebra brackets can be found in~\cite[p.~261]{Frappat}, \cite{Scheunert1977} or~\cite{JV2012}:
\begin{align}
& \{F^\pm,G^\pm\}=E^\pm, \quad \{F^\pm,G^\mp\}=Z\mp H, \nonumber\\
& \{F^\pm,F^\pm\}=\{G^\pm,G^\pm\}=\{F^\pm,F^\mp\}=\{G^\pm,G^\mp\}=0; \label{co-odd} \\
& [H,E^\pm]=\pm E^\pm,\quad [E^+,E^-]=2H,\quad [Z,H]=[Z,E^\pm]=0; \label{co-even} \\
&[H,F^\pm]=\pm\frac12 F^\pm, \quad [Z,F^\pm]=\frac12 F^\pm, \quad [E^\pm,F^\pm]=0,\quad [E^\mp,F^\pm]=-F^\mp, \nonumber\\
&[H,G^\pm]=\pm\frac12 G^\pm, \quad [Z,G^\pm]=-\frac12 G^\pm, \quad [E^\pm,G^\pm]=0,\quad [E^\mp,G^\pm]=G^\mp. \label{co-mixed}
\end{align}

The finite-dimensional irreducible representations of $\ssl(2|1)$ have been studied by 
Scheunert {\em et al}~\cite{Scheunert1977} and Marcu~\cite{Marcu}, but little seems to be known about
infinite-dimensional representations.
Here, we present a class of infinite-dimensional representations labeled by a positive number $\beta>0$.
We shall call them positive discrete series representations of $\ssl(2|1)$, as they are closely related
to positive discrete series representations of $\su(1,1)$.

First of all, let us fix a $\star$-structure (or an adjoint operation) on the Lie superalgebra by
\begin{equation}
Z^\dagger=Z,\quad H^\dagger=H,\quad (E^\pm)^\dagger = -E^\mp,\quad
(F^\pm)^\dagger = \mp G^\mp,\quad (G^\pm)^\dagger = \pm F^\mp.
\label{adjoint}
\end{equation}
It is easy to see that this $\star$-structure is indeed compatible with the 
Lie superalgebra brackets~\eqref{co-odd}-\eqref{co-mixed}.
Note that the $\star$-structure on $\ssl(2)$ implies that we are dealing with the form $\su(1,1)$.

The positive discrete series representations of $\ssl(2|1)$ are unitary representations
labeled by $\beta > 0$. The representation space is $\ell^2({\mathbb Z}_+)$ equipped with an orthonormal basis
$|\beta,n\rangle$ ($n=0,1,2,\ldots$), i.e.
\begin{equation}
\langle \beta,m | \beta,n \rangle = \delta_{m,n}.
\end{equation}
For the actions of the $\ssl(2|1)$ basis elements on these vectors, it is handy to
use the following ``even'' and ``odd'' functions, defined on integers $n$:
\begin{align}
& {\cal E}(n)=1 \hbox{ if } n \hbox{ is even and 0 otherwise},\nonumber\\
& {\cal O}(n)=1 \hbox{ if } n \hbox{ is odd and 0 otherwise}. \label{EO}
\end{align}
Note that ${\cal O}(n)=1-{\cal E}(n)$, but it is convenient to use both notations.
The actions of the odd generators are now given by:
\begin{align}
& F^+ |\beta,n\rangle = {\cal E}(n) \sqrt{\beta+\frac{n}{2}}\; |\beta,n+1\rangle, 
\qquad F^- |\beta,n\rangle = {\cal E}(n) \sqrt{\frac{n}{2}}\; |\beta,n-1\rangle, \nonumber\\
& G^+ |\beta,n\rangle = {\cal O}(n) \sqrt{\frac{n+1}{2}}\; |\beta,n+1\rangle, 
\qquad G^- |\beta,n\rangle = - {\cal O}(n) \sqrt{\beta+\frac{n-1}{2}}\; |\beta,n-1\rangle.
\label{FGact}
\end{align} 
The actions of the even generators can in principle be computed from~\eqref{co-odd}, and are
\begin{align}
& Z |\beta,n\rangle = -{\cal E}(n) \frac{\beta}{2}\; |\beta,n\rangle -
 {\cal O}(n) (\frac{\beta-1}{2})\; |\beta,n\rangle , \nonumber\\
& H |\beta,n\rangle = \frac{1}{2}(n+\beta)\; |\beta,n\rangle , \nonumber\\
& E^+ |\beta,n\rangle = {\cal E}(n) \sqrt{(\beta+\frac{n}{2})(1+\frac{n}{2})}\; |\beta,n+2\rangle 
 +{\cal O}(n) \sqrt{(\beta+\frac{n+1}{2})(\frac{n+1}{2})}\; |\beta,n+ 2\rangle, \nonumber \\
& E^- |\beta,n\rangle = -{\cal E}(n) \sqrt{(\beta-1+\frac{n}{2})(\frac{n}{2})}\; |\beta,n-2\rangle 
 -{\cal O}(n) \sqrt{(\beta+\frac{n-1}{2})(\frac{n-1}{2})}\; |\beta,n- 2\rangle .
\label{EHZact} 
\end{align}

\begin{prop}
The representation $\Pi_\beta$ ($\beta>0$) in $\ell^2({\mathbb Z}_+)$, defined by
the actions~\eqref{FGact}-\eqref{EHZact}, is an irreducible star representation (or unitary
representation) of the Lie superalgebra $\ssl(2|1)$.
With respect to the even subalgebra $\ssl(2)\cong \su(1,1)$, it decomposes into the direct sum
of two positive discrete series representations $\pi_{\frac{\beta}{2}}$ and $\pi_{\frac{\beta+1}{2}}$,
with Bargmann index (or label) resp.\ $\frac{\beta}{2}$ and $\frac{\beta+1}{2}$.
\end{prop}

\noindent {\bf Proof.}
To show that the actions~\eqref{FGact}-\eqref{EHZact} define indeed a representation of $\ssl(2|1)$ is
straightforward but tedious. Essentially, one should verify that all bracket relations~\eqref{co-odd}-\eqref{co-mixed}
are satisfied on the basis vectors $|\beta,n\rangle$. In practice, note that~\eqref{co-mixed} and~\eqref{co-even} follow
from~\eqref{co-odd} and the Jacobi-identity. So it is sufficient to verify that~\eqref{co-odd} is satisfied for
the action~\eqref{FGact}, and that~\eqref{FGact} implies~\eqref{EHZact} using~\eqref{co-odd}.

Note that $|\beta,0\rangle$ is a generating vector for the representation, since
\begin{equation}
(G^+ F^+)^n |\beta,0 \rangle = \sqrt{n!(\beta)_n}\;|\beta,2n\rangle, \qquad
F^+(G^+ F^+)^n |\beta,0 \rangle= \sqrt{n!(\beta)_{n+1}}\;|\beta,2n+1\rangle,
\end{equation}
where $(\beta)_n$ is the Pochhammer symbol~\cite{Andrews,Bailey,Slater}: $(\beta)_n=\beta(\beta+1)\cdots (\beta+n-1)$.
Irreducibility then follows from the actions
\begin{equation}
(G^- F^-)^n |\beta,2n\rangle = (-1)^n \sqrt{n!(\beta)_n}\;|\beta,0\rangle, \qquad
(G^- F^-)^n G^- |\beta,2n+1\rangle = (-1)^{n+1}\sqrt{n!(\beta)_{n+1}}\;|\beta,0\rangle.
\end{equation}

To see that the representation is a star representation for the $\star$-structure~\eqref{adjoint}, 
it is sufficient to check
\begin{align}
&\langle \beta,2n+1 | F^+ |\beta, 2n \rangle = - \langle \beta,2n | G^- |\beta, 2n+1 \rangle, \nonumber\\
&\langle \beta,2n-1 | F^- |\beta, 2n \rangle =  \langle \beta,2n | G^+ |\beta, 2n-1 \rangle.
\end{align}

Finally, relabel the even and odd vectors of the representation by $e_m=|\beta,2m\rangle$ and $f_m=|\beta,2m+1\rangle$.
Then the vectors $e_m$ ($m=0,1,2,\ldots$) are an orthonormal basis for the action of $\su(1,1)$ in $\ell^2({\mathbb Z}_+)$,
with 
\[
H e_m = (m+\frac{\beta}{2}) e_m, \quad E^+ e_m = \sqrt{(\beta+m)(m+1)}\, e_{m+1}, \quad
E^- e_m = -\sqrt{(\beta+m-1)m}\, e_{m-1},
\]
so this is the representation $\pi_{\frac{\beta}{2}}$. Note that the action
of $Z$ on this representation is $-\frac{\beta}{2}$ times the identity operator.
Similarly, the vectors $f_m$ ($m=0,1,2,\ldots$) are an orthonormal basis for the action of $\su(1,1)$ in $\ell^2({\mathbb Z}_+)$,
with in particular $H f_m = (m+\frac{\beta+1}{2}) f_m$, so this is the representation $\pi_{\frac{\beta+1}{2}}$. 
In this case, the action of $Z$ on this representation is $-\frac{\beta-1}{2}$ times the identity operator.
\mybox

\section{An $\ssl(2|1)$ oscillator model and the spectrum of a position operator}

In order to use the discrete series representations of $\ssl(2|1)$ for an oscillator model,
it is natural to take the Hamiltonian $\hat H$ as
\begin{equation}
\hat H = 2H +\frac12 -\beta.
\label{Ham}
\end{equation}
This operator is diagonal, self-adjoint, and has the equidistant spectrum:
$n+\frac12$ ($n=0,1,2,\ldots$).
Following the arguments of~\cite{JV2012}, we should take for the position operator $\hat q$
an arbitrary odd (real) self-adjoint element of $\ssl(2|1)$, i.e.\ an element of the form
\begin{equation}
\hat q =  F^+ + \gamma G^+ - G^- +\gamma F^-,
\label{hatq}
\end{equation}
with $\gamma$ a real constant (an overall constant does not play a crucial role, so 
that is why we have taken the coefficient of $F^+$ equal to 1, and \eqref{hatq} still represents the
most general case). 

When $\hat q$ is fixed by~\eqref{hatq}, the expression of the momentum operator $\hat p$ follows from~\eqref{Hqp}:
\begin{equation}
\hat p = \ii (F^+ + \gamma G^+ + G^- -\gamma F^-).
\label{hatp}
\end{equation}
These operators~\eqref{Ham}, \eqref{hatq} and~\eqref{hatp} do indeed satisfy~\eqref{Hqp} 
and the conditions described in section~1 are satisfied, and thus we are dealing with models for the 
oscillator in a class of infinite dimensional representations of $\ssl(2|1)$.

In the (ordered) basis $\{ |\beta,n\rangle, \; n=0,1,2,\ldots\}$, the operator $\hat q$ is
represented by an infinite symmetric tridiagonal matrix $M_q$:
\begin{equation}
M_q=\left(
\begin{array}{cccccc}
0 & R_0&  & & &  \\
R_0 & 0 & S_1 & & & \\
 & S_1 & 0 & R_1 & & \\
 & & R_1 & 0 & S_2 & \\
 & & & S_2 & 0 & \ddots \\
 & & & & \ddots & \ddots \\
\end{array}
\right) ,
\label{Mq}
\end{equation}
where
\begin{equation}
R_n=\sqrt{\beta+n}, \qquad S_n=\gamma\sqrt{n} \qquad (n=0,1,2,\ldots).
\label{RS}
\end{equation}
For $\gamma>0$, such a matrix is a Jacobi matrix, and its spectral theory is related to 
orthogonal polynomials~\cite{Berezanskii,Koelink1998,Koelink2004} ($\gamma<0$ is similar to $\gamma>0$: 
it will soon be clear that only $|\gamma|$ plays a role; for $\gamma=0$ the matrix~\eqref{Mq} decomposes and also
that case will be easy to treat).
Following the procedure described in~\cite[\S 2]{Koelink1998}, one should construct polynomials $p_n(x)$ of degree $n$
in $x$, with $p_{-1}(x)=0$, 
$p_0(x)=1$, and
\begin{align}
&x p_{2n}(x) = S_n p_{2n-1}(x) + R_n p_{2n+1}(x), \nonumber \\
&x p_{2n+1}(x) = R_n p_{2n}(x) + S_{n+1} p_{2n+2}(x), \qquad (n=0,1,2,\ldots).
\label{def-px}
\end{align}
Such polynomials are orthogonal for some positive weight function $w(x)$.
Then the spectrum of $M_q$ (or of $\hat q$) is the support of this weight function.
This technique works provided the (Hamburger) moment problem
for the Jacobi matrix is determinate~\cite{Berezanskii,Koelink2004}. This is equivalent to saying
that the corresponding Jacobi operator is essentially self-adjoint. A sufficient condition is 
that~\cite{Berezanskii,Koelink2004}
\[
\sum_{n=0}^{\infty} \frac{1}{R_n}+\sum_{n=1}^{\infty} \frac{1}{S_n}= 
\sum_{n=0}^{\infty} \frac{1}{\sqrt{\beta+n}}+\sum_{n=1}^{\infty} \frac{1}{\gamma\sqrt{n}}= \infty,
\]
which is satisfied here. So the spectrum of the position operator $\hat q$ is just the support of the weight function $w(x)$.
Furthermore, for a real value $x$ belonging to this support, the corresponding formal eigenvector of $\hat q$ is given by
\begin{equation}
v(x) = \sum_{n=0}^\infty p_n(x) \,|\beta,n\rangle.
\label{vx}
\end{equation}
So the purpose is first to construct the polynomials $p_n(x)$, and then to find the corresponding weight function.
The solution of~\eqref{def-px} is given in terms of terminating hypergeometric series; for their notation we 
follow that of standard books~\cite{Andrews,Bailey,Slater}.
\begin{prop}
\label{prop2}
When $\gamma^2\ne 1$, the solution of the recurrence relations~\eqref{def-px} is given by
\begin{align}
& p_{2n}(x) = (-\gamma)^{-n} \sqrt{\frac{(\beta)_n}{n!}}
 {\;}_2F_1\left( \myatop{-n,\frac{x^2}{1-\gamma^2}}{\beta} ; 1-\gamma^2 \right), \nonumber\\
& p_{2n+1}(x) = x (-\gamma)^{-n} \sqrt{\frac{(\beta+1)_n}{n!\beta}}
 {\;}_2F_1\left( \myatop{-n,\frac{x^2}{1-\gamma^2}+1}{\beta+1} ; 1-\gamma^2 \right). \label{2F1}
\end{align}
When $\gamma^2= 1$, the solution is of a different type and given by
\begin{align}
& p_{2n}(x) = (-\gamma)^{n} \sqrt{\frac{(\beta)_n}{n!}}
 {\;}_1F_1\left( \myatop{-n}{\beta} ; x^2 \right), \nonumber\\
& p_{2n+1}(x) = x (-\gamma)^{n} \sqrt{\frac{(\beta+1)_n}{n!\beta}}
 {\;}_1F_1\left( \myatop{-n}{\beta+1} ; x^2 \right).   \label{1F1}
\end{align}
\end{prop}
The proof is rather straightforward. It can be deduced from certain forward or backward shift operator formulas for 
the orthogonal polynomials that can be identified with the above expressions (see later). Alternatively, \eqref{2F1}
follows from the following contiguous relations for (terminating) hypergeometric series:
\begin{align}
& (b+n) {\;}_2F_1\left( \myatop{-n,a}{b+1} ; z \right) -n(1-z) {\;}_2F_1\left( \myatop{-n+1,a}{b+1} ; z \right)=
b {\;}_2F_1\left( \myatop{-n,a-1}{b} ; z \right), \nonumber\\
& {}_2F_1\left( \myatop{-n,a}{b} ; z \right) - {\;}_2F_1\left( \myatop{-n-1,a}{b} ; z \right)=
\frac{az}{b} {\ }_2F_1\left( \myatop{-n,a+1}{b+1} ; z \right). \label{contig1}
\end{align}
Such contiguous relations are trivial to verify by comparing coefficients of $z$ in left and right hand side.
Similarly, \eqref{1F1} follows from:
\begin{align}
& (b+n) {\;}_1F_1\left( \myatop{-n}{b+1} ; z \right) -n {\ }_1F_1\left( \myatop{-n+1}{b+1} ; z \right)=
b {\;}_1F_1\left( \myatop{-n}{b} ; z \right), \nonumber\\
& {}_1F_1\left( \myatop{-n}{b} ; z \right) - {\;}_1F_1\left( \myatop{-n-1}{b} ; z \right)=
\frac{z}{b} {\ }_1F_1\left( \myatop{-n}{b+1} ; z \right). \label{contig2}
\end{align} 
 
Now it is a matter of identifying the above polynomials in order to find $w(x)$.
For this purpose, recall the definition of the Meixner polynomial $M_n(k;\beta,c)$ of degree $n$ in $k$, with
parameters $\beta$ and $c$~\cite{Koekoek,Ismail,Andrews}:
\begin{equation}
M_n(k;\beta,c) = {\ }_2F_1\left( \myatop{-n,-k}{\beta} ; 1-\frac{1}{c} \right). \label{Meixner}
\end{equation}
These polynomials satisfy a discrete orthogonality relation:
\begin{equation}
\sum_{k=0}^{\infty} \frac{(\beta)_k}{k!} c^k M_m(k;\beta,c) M_n(k;\beta,c) = \frac{c^{-n}n!}{(\beta)_n(1-c)^\beta}
\delta_{mn}
\label{M-orth}
\end{equation}
when $\beta>0$ and $0<c<1$.

Whether the polynomials in Proposition~\ref{prop2} can be identified with Meixner polynomials, depends on $\gamma$. 
We should distinguish four cases.

\vskip 1mm
\noindent {\bf Case 1: $|\gamma|>1$.}

From~\eqref{2F1} it is clear that we can identify $c$ with $1/\gamma^2$, and we have
\begin{align}
& p_{2n}(x) = (-\gamma)^{-n} \sqrt{\frac{(\beta)_n}{n!}} M_n(\frac{x^2}{\gamma^2-1}; \beta,\frac{1}{\gamma^2}), \nonumber\\
& p_{2n+1}(x) = x (-\gamma)^{-n} \sqrt{\frac{(\beta+1)_n}{n!\beta}} M_n(\frac{x^2}{\gamma^2-1}-1; \beta+1,\frac{1}{\gamma^2}). 
\label{Mcase1}
\end{align}
The orthogonality relation~\eqref{M-orth} leads to the following result:
\begin{prop}
\label{prop3}
For $|\gamma|>1$, the polynomials $p_n(x)$ satisfy a discrete orthogonality relation:
\begin{equation}
\sum_{x\in {\cal S}_1} w(x) p_n(x)p_m(x) = \left(\frac{\gamma^2}{\gamma^2-1} \right)^\beta \delta_{mn}, \label{ort1}
\end{equation}
where
\begin{equation}
{\cal S}_1=\{ \pm \sqrt{\gamma^2-1} \sqrt{k}  \ |\  k \in {\mathbb Z}_+ \}, \label{S1}
\end{equation}
and where the weight function is given by 
\begin{align}
w(x) &= 1 \qquad \hbox{for } x=0, \nonumber\\
w(x) &= \frac{1}{2} \frac{(\beta)_k}{k!} \gamma^{-2k} \qquad \hbox{for } x=\pm \sqrt{\gamma^2-1} \sqrt{k} \quad(k=1,2,\ldots).
\label{w1}
\end{align}
\end{prop} 
So in this case, the spectrum of the position operator $\hat q$ is discrete and given by~\eqref{S1}.

\vskip 1mm
\noindent {\bf Case 2: $|\gamma|=1$.}

For $|\gamma|=1$, we have already deduced in Proposition~\ref{prop2} that the polynomials are ${}_1F_1$ series, and
these can be identified with (generalized) Laguerre polynomials.
\begin{equation}
p_{2n}(x) = (-\gamma)^{n} \sqrt{\frac{n!}{(\beta)_n}} L_n^{(\beta-1)}(x^2),\qquad 
p_{2n+1}(x) = (-\gamma)^{n} \sqrt{\frac{n!}{(\beta)_{n+1}}}\; x L_n^{(\beta)}(x^2).   \label{Lcase2}
\end{equation}
The orthogonality relation of Laguerre polynomials leads to the following result:
\begin{prop}
\label{prop4}
For $|\gamma|=1$, the polynomials $p_n(x)$ satisfy a continuous orthogonality relation:
\begin{equation}
\int_{-\infty}^{+\infty} w(x) p_n(x)p_m(x)dx = \Gamma(\beta) \delta_{mn}, \label{ort2}
\end{equation}
where
\begin{equation}
w(x) = \ee^{-x^2} |x|^{2\beta-1}.
\label{w2}
\end{equation}
\end{prop} 
So in the second case, the spectrum of the position operator $\hat q$ is continuous and given by ${\mathbb R}$.

\vskip 1mm
\noindent {\bf Case 3: $0<|\gamma|<1$.}

Having found Meixner and Laguerre polynomials for the first and second case, one might expect to find Meixner-Pollaczek
polynomials for the third case. However, this is not so. 
To see the proper form, one should first apply a transformation on the ${}_2F_1$ series in~\eqref{2F1},
\begin{equation}
{}_2F_1\left( \myatop{-n,a}{b} ; z \right) = (1-z)^{-n} {\ }_2F_1\left( \myatop{-n,b-a}{b} ; \frac{z}{z-1} \right).
\label{tf}
\end{equation}
After this, one can again identify the polynomials with Meixner polynomials (now with $c=\gamma^2$):
\begin{align}
& p_{2n}(x) = (-\gamma)^{n} \sqrt{\frac{(\beta)_n}{n!}} M_n(\frac{x^2}{1-\gamma^2}-\beta; \beta,\gamma^2), \nonumber\\
& p_{2n+1}(x) = x (-\gamma)^{n} \sqrt{\frac{(\beta+1)_n}{n!\beta}} M_n(\frac{x^2}{1-\gamma^2}-\beta; \beta+1,\gamma^2). 
\label{Mcase3}
\end{align}
The orthogonality relation~\eqref{M-orth} now leads to the following result:
\begin{prop}
\label{prop5}
For $0<|\gamma|<1$, the polynomials $p_n(x)$ satisfy a discrete orthogonality relation:
\begin{equation}
\sum_{x\in {\cal S}_3} w(x) p_n(x)p_m(x) = \frac{1}{(1-\gamma^2)^\beta} \delta_{mn}, \label{ort3}
\end{equation}
where
\begin{equation}
{\cal S}_3=\{ \pm \sqrt{1-\gamma^2} \sqrt{\beta+k}  \ |\  k \in {\mathbb Z}_+ \}, \label{S3}
\end{equation}
and where the weight function is given by 
\begin{equation}
w(x) = \frac{1}{2} \frac{(\beta)_k}{k!} \gamma^{2k} \qquad \hbox{for } x=\pm \sqrt{1-\gamma^2} \sqrt{\beta+k} \quad(k=0,1,2,\ldots).
\label{w3}
\end{equation}
\end{prop} 
So in the third case, the spectrum of the position operator $\hat q$ is again discrete and given by~\eqref{S3}.

\vskip 1mm
\noindent {\bf Case 4: $\gamma=0$.}

In principle, there is a fourth case with $\gamma=0$, but this is somewhat trivial and we shall not return to it later.
Indeed, for $\gamma=0$ the matrix $M_q$ falls apart into irreducible $(2\times 2)$-blocks, because all $S_n=0$.
The spectrum corresponds to the eigenvalues of these $(2\times 2)$-blocks, and it turns out 
that the corresponding polynomials are just discrete delta-functions:
\begin{prop}
\label{prop6}
For $\gamma=0$, the polynomials $p_n(x)$ satisfy the discrete orthogonality relation:
\begin{equation}
\sum_{x\in {\cal S}_4} w(x) p_n(x)p_m(x) = \delta_{mn}, \label{ort4}
\end{equation}
where
\begin{equation}
{\cal S}_4=\{ \pm \sqrt{\beta+k}  \ |\  k \in {\mathbb Z}_+ \}, \label{S4}
\end{equation}
and where the weight function is constant:
\begin{equation}
w(x) = \frac{1}{2} \qquad \hbox{for } x=\pm \sqrt{\beta+k} \quad(k=0,1,2,\ldots).
\label{w4}
\end{equation}
The polynomials are 
\begin{align}
& p_{2n}(x) = \delta_{n,k} \qquad \hbox{for } x=\pm\sqrt{\beta+k}, \nonumber\\
& p_{2n+1}(x) = \pm \delta_{n,k} \qquad \hbox{for } x=\pm\sqrt{\beta+k}. 
\end{align}
\end{prop} 

The above analysis determines the spectrum of the position operator $\hat q$ in the representation determined by $\beta>0$,
and for all possible values of the parameter $\gamma$ in~\eqref{hatq}. 
For each $x$ belonging to the spectrum of $\hat q$, the corresponding formal eigenvector is given by~\eqref{vx}.
Since essentially only $\gamma^2$ plays a role, and since $\gamma=0$ is a redundant case, 
we shall in the rest of the paper deal with $\gamma>0$.

Now the determination of the spectrum and eigenvectors of the momentum operator $\hat p$ is a formality.
Due to the simple connection between~\eqref{hatq} and~\eqref{hatp}, one can easily deduce that the spectrum of $\hat p$
is the same as that of $\hat q$ (in all four cases). Furthermore, the formal eigenvector of $\hat p$ for the eigenvalue $y$
is given by
\begin{equation}
w(y) = \sum_{n=0}^\infty (-\ii)^n p_n(y) \,|\beta,n\rangle,
\label{wy}
\end{equation}
where the $p_n$'s are the same polynomials that appear in the analysis of $\hat q$.

\section{On the shape of position and momentum wavefunctions}

The position (resp.\ momentum) wavefunctions of the $\ssl(2|1)$ finite oscillator are the overlaps 
between the (normalized) $\hat q$-eigenvectors~\eqref{vx} (resp.\ $\hat p$-eigenvectors~\eqref{wy})
and the $\hat H$-eigenvectors. 
Because of the close relation between~\eqref{vx} and~\eqref{wy} it will be sufficient to study 
only the position wavefunctions.
Obviously, these wavefunctions depend on the representation parameter $\beta>0$, and on the parameter $\gamma$ 
which appears in the expression~\eqref{hatq} of $\hat q$, so we will denote them by $\Phi_n^{(\beta,\gamma)}(x)$:
\begin{equation}
v(x) = \sum_{n=0}^\infty \Phi_n^{(\beta,\gamma)}(x)\; |\beta,n\rangle.
\label{Phi}
\end{equation}
Herein, $x$ belongs to the spectrum of $\hat q$, and $\Phi_n^{(\beta,\gamma)}(x)$ is the polynomial $p_n(x)$ 
as in~\eqref{vx} but normalized.

Let us begin with a familiar case, namely for $\gamma=1$, when the spectrum of $\hat q$ is ${\mathbb R}$. 
From~\eqref{ort2} and \eqref{w2}, it follows that the normalized versions of~\eqref{Lcase2} are given by
\begin{align}
\Phi_{2n}^{(\beta,1)}(x) & = (-1)^n  \sqrt {\frac{n!}{\Gamma( n + \beta) } }\, 
	|x|^{\beta-1/2}\,  \ee^{-x^2/2} L_n^{(\beta-1)}(x^2), \nonumber \\
\Phi_{2n+1}^{(\beta,1)}(x) & = (-1)^n  \sqrt {\frac{n!}{\Gamma( n + \beta+1) } }\,  
|x|^{\beta-1/2}\, \ee^{-x^2/2} x L_n^{(\beta)}(x^2).
\label{wave2}
\end{align}
These are just the paraboson wavefunctions, see e.g.~\cite[(A.11)]{JSV2011} (with $\beta$ equal to the paraboson
parameter $a$ in~\cite{JSV2011}).
In particular, when the paraboson parameter equals $1/2$, one is just in the ordinary boson case; and indeed
one has:
\begin{equation}
\Phi_n^{(1/2,1)}(x) = \frac{1}{2^{n/2}\pi^{1/4}\sqrt{n!}} \ee^{-x^2/2} H_n(x),
\end{equation}
with $H_n(x)$ the common Hermite polynomial. 
So for $(\beta,\gamma)=(1/2,1)$, the $\ssl(2|1)$ oscillator coincides with the canonical oscillator, whereas for
$(\beta,\gamma)=(\beta,1)$ it coincides with the paraboson oscillator with paraboson parameter $\beta$.

Let us now consider the cases $0<\gamma<1$ and $\gamma>1$.
For $0<\gamma<1$, the expressions of the (discrete) wavefunctions follow from~\eqref{Mcase3}, using~\eqref{w3} and the
normalization~\eqref{ort3}:
\begin{align}
\Phi_{2n}^{(\beta,\gamma)}(x) & = (-1)^n \gamma^{n+k}\sqrt{\frac{(\beta)_n(\beta)_k}{2n!k!}} (1-\gamma^2)^{\beta/2} M_n(k;\beta,\gamma^2), \nonumber\\
\Phi_{2n+1}^{(\beta,\gamma)}(x) & = (-1)^n \gamma^{n+k}\sqrt{\frac{(\beta+1)_n(\beta)_k}{2\beta n!k!}} (1-\gamma^2)^{\beta/2} x M_n(k;\beta+1,\gamma^2), \label{wave3} \\
&\qquad\hbox{where } x=\pm\sqrt{1-\gamma^2}\sqrt{\beta+k},\qquad (k=0,1,2,\ldots). \nonumber
\end{align}
In a similar way, one obtains for $\gamma>1$:
\begin{align}
\Phi_{2n}^{(\beta,\gamma)}(x) & = (-1)^n \gamma^{-n-k-\beta}
\sqrt{\frac{(\beta)_n(\beta)_k}{2n!k!}} (\gamma^2-1)^{\beta/2} M_n(k;\beta,\frac{1}{\gamma^2}), \nonumber\\
\Phi_{2n+1}^{(\beta,\gamma)}(x) & = (-1)^n \gamma^{-n-k-\beta}
\sqrt{\frac{(\beta+1)_n(\beta)_k}{2\beta n!k!}} (\gamma^2-1)^{\beta/2} x M_n(k-1;\beta+1,\frac{1}{\gamma^2}), \label{wave1} \\
&\qquad\hbox{where } x=\pm\sqrt{\gamma^2-1}\sqrt{k}, \qquad (k=1,2,\ldots); \nonumber
\end{align}
for $x=0$, the weight function is not simply the last expression in~\eqref{w1}, and therefore we have a separate
expression for the wavefunction at $x=0$:
\begin{equation}
\Phi_{2n}^{(\beta,\gamma)}(0)  = (-1)^n \gamma^{-n-\beta}
\sqrt{\frac{(\beta)_n}{n!}} (\gamma^2-1)^{\beta/2}, \qquad \Phi_{2n+1}^{(\beta,\gamma)}(0)  =0.
\label{wave10}
\end{equation}
Note that we could also keep the first expression in~\eqref{wave1} and multiply it by $\sqrt{1+\delta_{k,0}}$;
then it coincides with~\eqref{wave10} for $k=0$ (or $x=0$), and then we have a unified expression for 
all $k=0,1,2,\ldots$.

\begin{figure}[th]
\[
\begin{tabular}{ccc}
\hline\\[-3mm]
 & $n=0$ & $n=1$ \\[1mm]
$\gamma=0.4$&\includegraphics[scale=0.70]{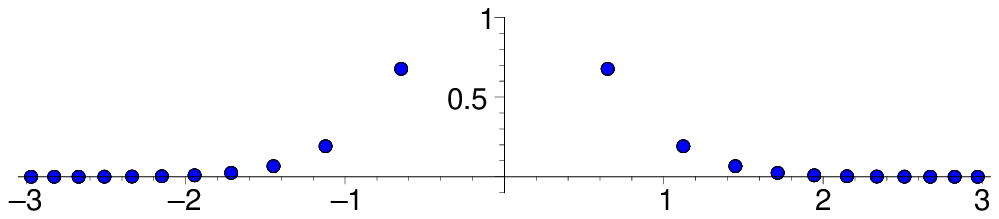} & \includegraphics[scale=0.70]{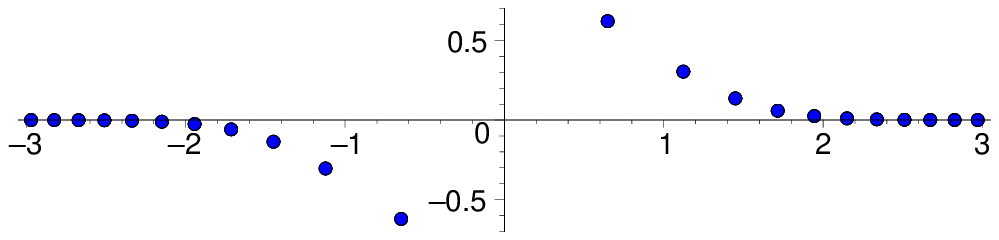} \\[1mm]
$\gamma=0.75$&\includegraphics[scale=0.70]{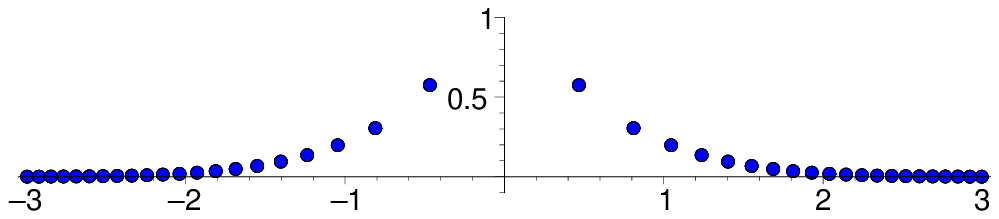} & \includegraphics[scale=0.70]{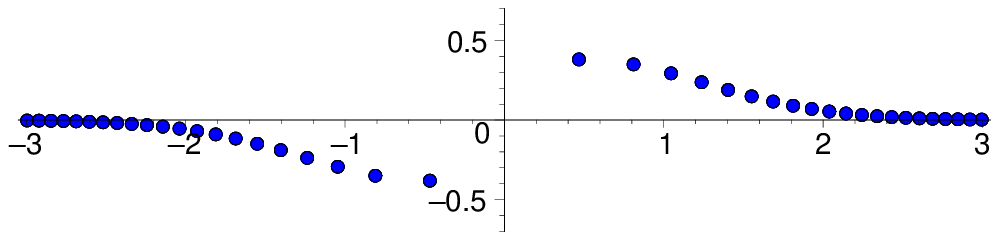} \\[1mm]
$\gamma=1$&\includegraphics[scale=0.70]{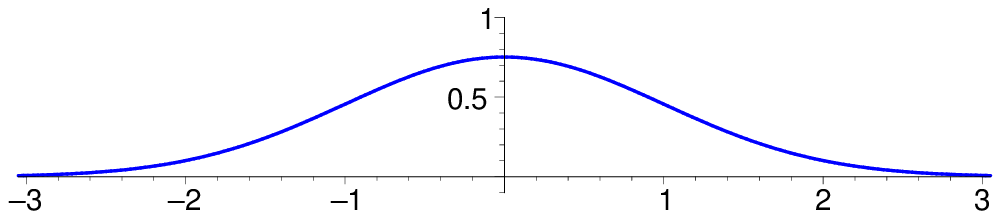} & \includegraphics[scale=0.70]{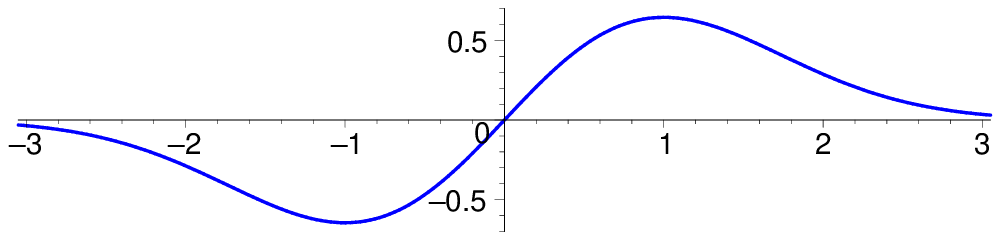} \\[1mm]
$\gamma=1.2$&\includegraphics[scale=0.70]{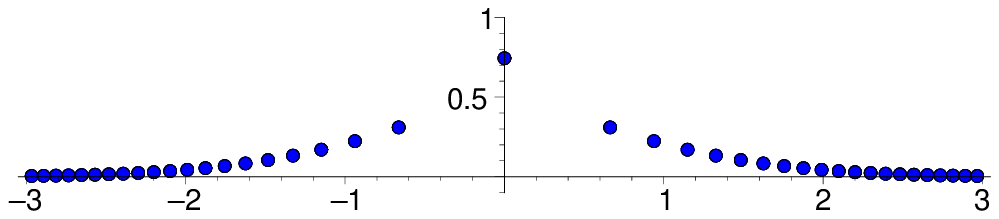} & \includegraphics[scale=0.70]{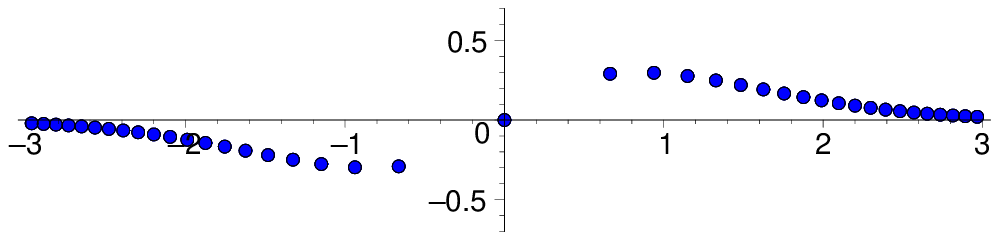} \\[1mm]
$\gamma=1.5$&\includegraphics[scale=0.70]{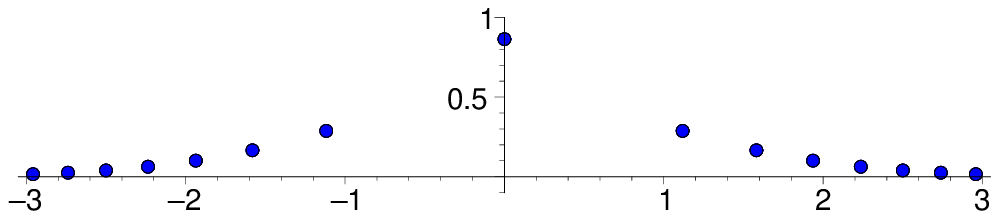} & \includegraphics[scale=0.70]{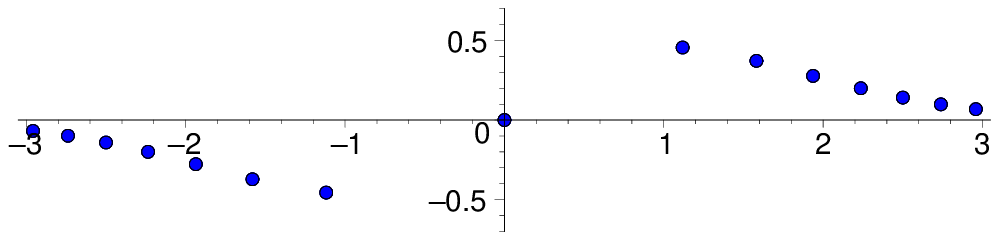} \\
\hline
\end{tabular} 
\]
\caption{Plots of the wavefunctions $\Phi^{(\beta,\gamma)}_n(x)$ in the representation with $\beta=1/2$,
for $n=0$ (left column), and $n=1$ (right column), for $\gamma=0.4, 0.75, 1.0, 1.2, 1.5$. For $\gamma=1$ the 
wavefunction is continuous, for the other $\gamma$-values it is discrete.}
\label{fig1}
\end{figure}

Let us now consider the plots of these wavefunctions $\Phi_{n}^{(\beta,\gamma)}(x)$, for some values of the parameters
$\beta$ and $\gamma$ and for some $n$-values.
As a first case, it is interesting to take $\beta=1/2$, since we know that the case $(\beta,\gamma)=(1/2,1)$ coincides
with the canonical quantum oscillator. 
In Figure~\ref{fig1} we have plotted the `ground state' wavefunction $\Phi_{0}^{(1/2,\gamma)}(x)$ and the 
`first excited state' wavefunction $\Phi_{1}^{(1/2,\gamma)}(x)$, for some values of $\gamma$. For $\gamma=1$, this yields
the common wavefunctions of the canonical oscillator, with support ${\mathbb R}$.
Then, we have plotted the wavefunctions for some values of $\gamma<1$, where the expression~\eqref{wave3} is used;
and for some values of $\gamma>1$, where the expression~\eqref{wave1} is used.
In both of these cases, the support of the wavefunction is discrete (but infinite), so the plots consist of an
infinite number of dots (of course, in the figures we can show only a finite number). 
Observe the similarity between the discrete plots of $\Phi_{n}^{(1/2,\gamma)}(x)$ for $\gamma\ne 1$ and the
continuous wavefunction $\Phi_{n}^{(1/2,1)}(x)$. When $\gamma$ tends to 1 (either from above or from below), 
the discrete plots of $\Phi_{n}^{(1/2,\gamma)}(x)$ tend to the continuous plot of $\Phi_{n}^{(1/2,1)}(x)$,
provided the dots in the discrete plot are properly redistributed. This is necessary because in the
continuous case the function satisfies
\[
\int_{-\infty}^{+\infty} \left(\Phi_{n}^{(1/2,1)}(x) \right)^2 dx =1,
\]
whereas in the discrete case one has
\[
\sum_{x\in{\cal S}} \left(\Phi_{n}^{(1/2,\gamma)}(x) \right)^2 =1 \qquad (\gamma\ne 1),
\]
with ${\cal S}$ the support given by \eqref{S1} or \eqref{S3}.
This observation also follows from the limits determined at the end of this section.

\begin{figure}[th]
\[
\begin{tabular}{ccc}
\hline\\[-3mm]
 & $n=0$ & $n=1$ \\[1mm]
$\gamma=0.4$&\includegraphics[scale=0.70]{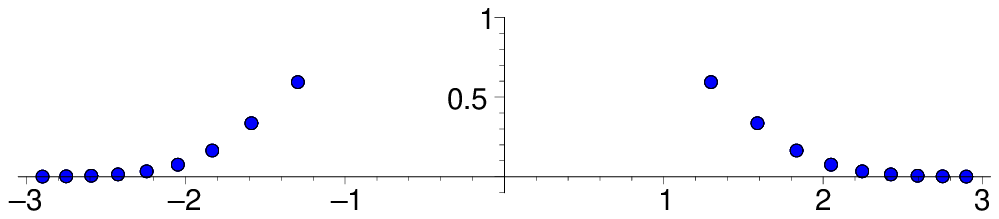} & \includegraphics[scale=0.70]{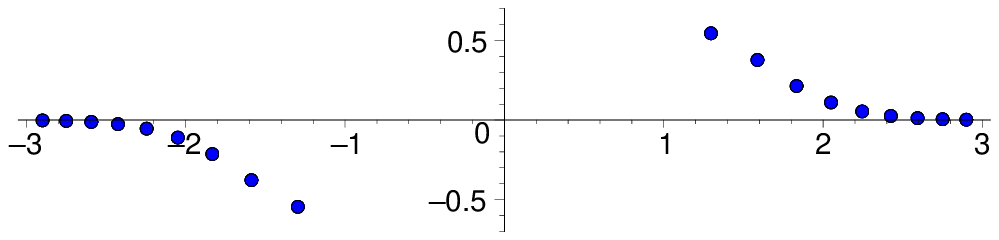} \\[1mm]
$\gamma=0.75$&\includegraphics[scale=0.70]{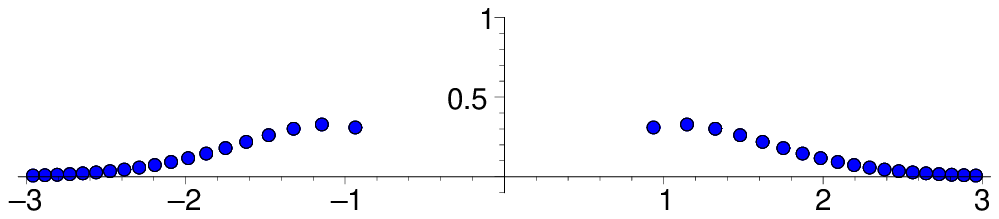} & \includegraphics[scale=0.70]{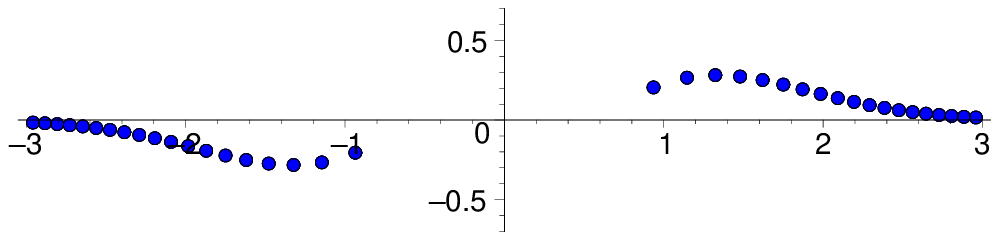} \\[1mm]
$\gamma=1$&\includegraphics[scale=0.70]{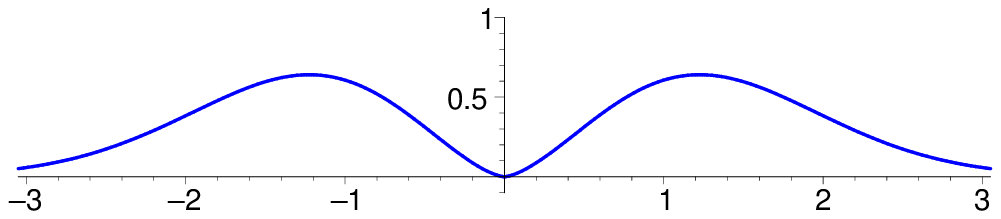} & \includegraphics[scale=0.70]{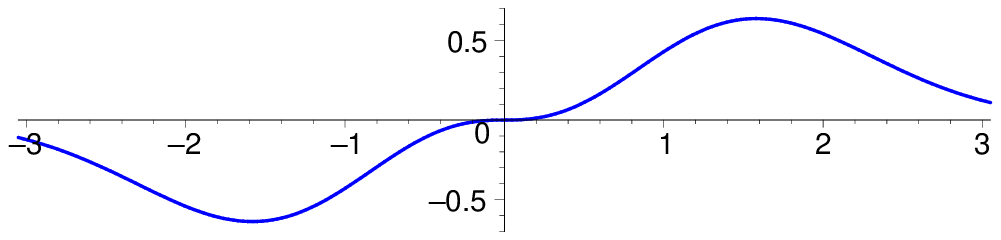} \\[1mm]
$\gamma=1.2$&\includegraphics[scale=0.70]{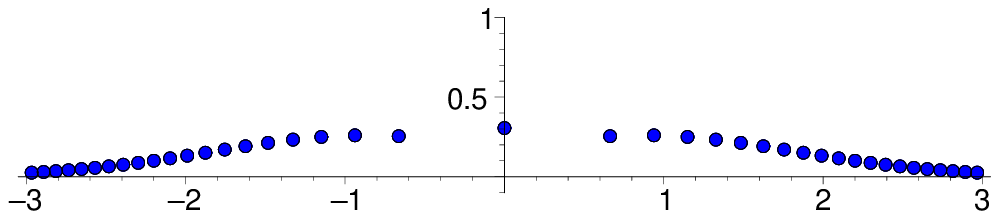} & \includegraphics[scale=0.70]{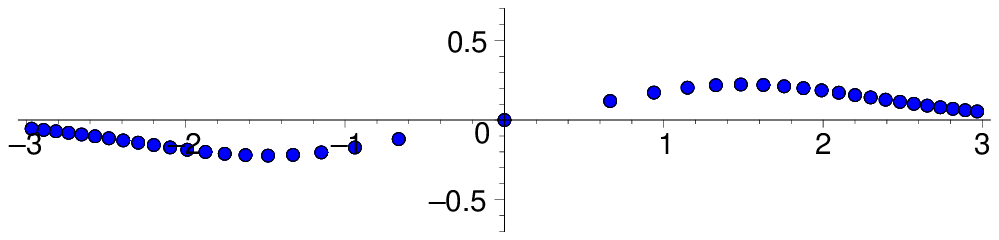} \\[1mm]
$\gamma=1.5$&\includegraphics[scale=0.70]{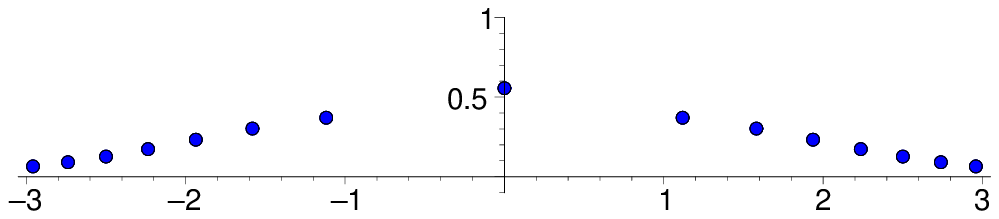} & \includegraphics[scale=0.70]{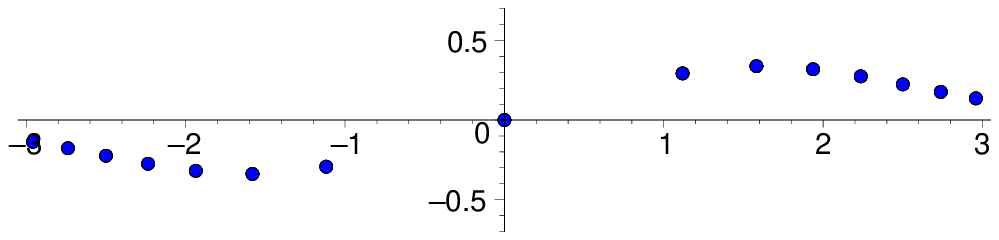} \\
\hline
\end{tabular} 
\]
\caption{Plots of the wavefunctions $\Phi^{(\beta,\gamma)}_n(x)$ in the representation with $\beta=2$,
for $n=0$ (left column), and $n=1$ (right column), for $\gamma=0.4, 0.75, 1.0, 1.2, 1.5$. For $\gamma=1$ the 
wavefunction is continuous, for the other $\gamma$-values it is discrete.}
\label{fig2}
\end{figure}

In Figure~\ref{fig2} we have plotted the `ground state' wavefunction $\Phi_{0}^{(\beta,\gamma)}(x)$ and the 
`first excited state' wavefunction $\Phi_{1}^{(\beta,\gamma)}(x)$, again for some values of $\gamma$, but now for
another value of $\beta$: $\beta=2$. For $\gamma=1$, this yields
the wavefunctions of the paraboson oscillator, with support ${\mathbb R}$.
The wavefunctions for the other values of $\gamma$ yield discrete plots that tend to the paraboson wavefunctions
when $\gamma$ tends to 1.

Let us briefly return to the limits of the discrete wavefunctions when $\gamma\rightarrow 1$.
In the case $0<\gamma<1$, consider the even wavefunction in~\eqref{wave3}. The essential limit comes from
\begin{equation}
\lim_{\gamma\rightarrow 1} M_n(k;\beta,\gamma^2) =
\lim_{\gamma\rightarrow 1} M_n(\frac{x^2}{1-\gamma^2}-\beta;\beta,\gamma^2) = \frac{n!}{(\beta)_n} L_n^{(\beta-1)}(x^2),
\end{equation}
which is a slightly modified form of a known limit~\cite[p.~243]{Koekoek}.
For the odd wavefunctions, the limit is similar, and also the factor coming from the weight function is
easily computed under the limit, so one finds indeed that the expressions~\eqref{wave3} yield those of~\eqref{wave2}
in the limit $\gamma\rightarrow 1$ ($0<\gamma<1$):
\[
\lim_{\gamma\rightarrow 1} \Phi_{n}^{(\beta,\gamma)}(x) = \Phi_{n}^{(\beta,1)}(x).
\]
The limit $\gamma\rightarrow 1$ for the case that $\gamma>1$, i.e.\ the wavefunctions~\eqref{wave1}, is the same and
the computation is similar to the one just described.

In this section, we have paid attention only to the position wavefunctions. The momentum wavefunctions are
completely analogous, and in fact it follows from~\eqref{wy} that they are given by
\begin{equation}
\Psi_n^{(\beta,\gamma)}(y) = (-\ii)^n \Phi_n^{(\beta,\gamma)}(y),
\label{Psi}
\end{equation}
where the last expression is that of the position wavefunction.

\section{Expressions for the $\ssl(2|1)$ Fourier transform}

In canonical quantum mechanics, the momentum wavefunction (in $L^2({\mathbb R})$) is given by the Fourier transform of
the position wavefunction (and vice versa), with kernel $K(x,y)$:
\[
\Psi(y)=  \int K(x,y) \Phi(x)dx, \qquad K(x,y)=\frac{1}{\sqrt{2\pi}} \ee^{-\ii xy}.
\]
This means that $K(x,y)$ is also the overlap of the (formal) position eigenvector for the eigenvalue $x$ with the
momentum eigenvector for the eigenvalue $y$.

So in the current case, the kernel for the corresponding $\ssl(2|1)$ Fourier transform is given by
\begin{equation}
K^{(\beta,\gamma)}(x,y) = \langle v(x), w(y)\rangle = \sum_{n=0}^\infty \Phi^{(\beta,\gamma)}_n(x) \Psi^{(\beta,\gamma)}_n(y) 
= \sum_{n=0}^\infty (-\ii)^n \Phi^{(\beta,\gamma)}_n(x) \Phi^{(\beta,\gamma)}_n(y).
\label{K}
\end{equation}
We need to compute this function in three cases, according to $0<\gamma<1$, $\gamma=1$ or $\gamma>1$.
Let us start with the known case $\gamma=1$. The corresponding kernel has been computed in a number
of papers~\cite{Mukunda,Regniers2010}:
\begin{equation}
K^{(\beta,1)}(x,y)= \frac{|xy|^{\beta-1/2}}{2^\beta \Gamma(\beta)} \left[
{\;}_0F_1\left( \myatop{-}{\beta} ; -\frac{x^2y^2}{4} \right) - \ii \frac{xy}{2\beta} 
{\;}_0F_1\left( \myatop{-}{\beta+1} ; -\frac{x^2y^2}{4} \right)
\right].
\label{F2}
\end{equation}
It is easy to see that for $\beta=1/2$ this expression reduces to
\[
\frac{1}{\sqrt{2\pi}}(\cos(xy)-\ii \sin(xy))= \frac{1}{\sqrt{2\pi}} \ee^{-\ii xy}.
\]

For the remaining cases, one can make use of the bilinear generating function for Meixner polynomials~\cite{AJNW1998}
(which can be deduced from~\cite[eq.~(12), p.~85]{Erdelyi} or~\cite[proposition~3]{Jagan1998}):
\begin{equation}
\sum_{n=0}^\infty \frac{(b)_n}{n!} z^n M_n(\xi;b,c) M_n(\xi';b,c)= (1-z)^{-b-\xi-\xi'} (1-\frac{z}{c})^{\xi+\xi'}
{\;}_2F_1\left( \myatop{-\xi,-\xi'}{b} ; \frac{z(1-c)^2}{(z-c)^2} \right).
\end{equation}
Using this, we obtained for the case $0<\gamma<1$:
\begin{align}
& K^{(\beta,\gamma)}(x,y)= \frac12 \left(\frac{2\gamma}{1+\gamma^2}\right)^{k+l} \left(\frac{1-\gamma^2}{1+\gamma^2}\right)^{\beta} 
\sqrt{\frac{(\beta)_k(\beta)_l}{k!l!}} \nonumber\\
&\times \left[ 
{\;}_2F_1\left( \myatop{-k,-l}{\beta} ; -\frac{1}{4}(\gamma-\frac{1}{\gamma})^2 \right) - \ii \frac{xy}{\beta(1+\gamma^2)} 
{\;}_2F_1\left( \myatop{-k,-l}{\beta+1} ; -\frac{1}{4}(\gamma-\frac{1}{\gamma})^2 \right)
\right]
\label{F3}, \\
&\hbox{where } x=\pm\sqrt{1-\gamma^2}\sqrt{\beta+k}, \quad y=\pm\sqrt{1-\gamma^2}\sqrt{\beta+l}, \quad (k,l=0,1,2,\ldots). \nonumber
\end{align}
In a similar way, one obtains for $\gamma>1$:
\begin{align}
& K^{(\beta,\gamma)}(x,y)= \frac12 \left(\frac{2\gamma}{1+\gamma^2}\right)^{k+l} \left(\frac{\gamma^2-1}{\gamma^2+1}\right)^{\beta} 
\sqrt{\frac{(\beta)_k(\beta)_l}{k!l!}} \sqrt{(1+\delta_{k,0})(1+\delta_{l,0})}\nonumber\\
&\times \left[ 
{\;}_2F_1\left( \myatop{-k,-l}{\beta} ; -\frac{1}{4}(\gamma-\frac{1}{\gamma})^2 \right) - \ii \frac{xy(1+\gamma^2)}{4\beta\gamma^2} 
{\;}_2F_1\left( \myatop{-k+1,-l+1}{\beta+1} ; -\frac{1}{4}(\gamma-\frac{1}{\gamma})^2 \right)
\right]
\label{F1}, \\
&\hbox{where } x=\pm\sqrt{\gamma^2-1}\sqrt{k}, \quad y=\pm\sqrt{\gamma^2-1}\sqrt{l}, \quad (k,l=0,1,2,\ldots). \nonumber
\end{align}

Note that one can explicitly compute the limits $\gamma\rightarrow 1$ of the kernels~\eqref{F3} and~\eqref{F1}, and
show that they yield the known paraboson kernel~\eqref{F2}.
Such computations are based on limit relations of the following type:
\begin{align*}
& \lim_{\myatop{\scriptstyle \gamma\rightarrow 1}{\scriptstyle \gamma<1}} 
 {\;}_2F_1\left( \myatop{-k,-l}{\beta} ; -\frac{1}{4}(\gamma-\frac{1}{\gamma})^2 \right) =
 \lim_{\myatop{\scriptstyle \gamma\rightarrow 1}{\scriptstyle \gamma<1}} 
 {\;}_2F_1\left( \myatop{\beta-\frac{x^2}{1-\gamma^2},\beta-\frac{y^2}{1-\gamma^2}}{\beta} ;
  -\frac{1}{4}(\gamma-\frac{1}{\gamma})^2 \right) \\
& = {\;}_0F_1\left( \myatop{-}{\beta} ; -\frac{x^2y^2}{4} \right).
\end{align*}

\section{The paraboson oscillator and $\osp(1|2)\subset \ssl(2|1)$}

Wigner~\cite{Wigner} introduced the one-dimensional Wigner oscillator or paraboson oscillator~\cite{Ohnuki}, 
leading to the field of Wigner quantization~\cite{Palev79,King2003}.
We shall recall some formulas for the paraboson oscillator (see~\cite{JLV2008} or the appendix of~\cite{JSV2011}).
In terms of the momentum and position operator $\hat p$ and $\hat q$, the Hamiltonian of the 
paraboson oscillator is given by 
\begin{equation}
\label{H0}
\hat H_0 = \frac{\hat p^2}{2} + \frac{\hat q^2}{2}.
\end{equation}
Wigner dropped the canonical commutation relation $[\hat q,\hat p] = \ii$,
but required instead the compatibility
between the Hamilton and the Heisenberg equations.
These compatibility conditions are:
\begin{equation}
\label{CCs}
[ \frac{\hat p^2}{2} + \frac{\hat q^2}{2}, \hat p] = \ii\hat q,\quad
[ \frac{\hat p^2}{2} + \frac{\hat q^2}{2}, \hat q] = -\ii\hat p.
\end{equation}
So, one has to find self-adjoint operators $\hat p$ and $\hat q$, acting in some Hilbert space,
such that the compatibility conditions~\eqref{CCs} hold.
The solutions to~\eqref{CCs} can be found by introducing two new
operators $b^+$ and $b^-$ (the paraboson creation and annihilation operators):
\begin{equation}\label{bpm}
b^\pm = \frac{1}{\sqrt{2}}(\hat q \mp \ii\hat p), 
\end{equation}
or equivalently
\begin{equation*}
\hat q = \frac{1}{\sqrt{2}}(b^+ + b^-),\quad
\hat p = \frac{\ii}{\sqrt{2}}(b^+ - b^-).
\end{equation*}
It is then easy to see that $\hat H_0 = \frac12\{b^-,b^+\}$,
and that the compatibility conditions~\eqref{CCs} are equivalent with
\begin{equation}
	[\{b^-,b^+\}, b^\pm] = \pm 2 b^\pm.
	\label{CCs2}
\end{equation}

The relations~\eqref{CCs2} are nowadays the defining relations of 
a pair of paraboson operators $b^\pm$~\cite{Green53}.
Furthermore, it is known that the Lie superalgebra generated by two odd elements 
$b^\pm$ subject to the restriction~\eqref{CCs2} is the Lie superalgebra
$\osp(1|2)$~\cite{Ganchev}.  
Keeping in mind the self-adjointness of $\hat q$ and $\hat p$, i.e.\ $(b^\pm)^\dagger = b^\mp$,
one is then faced with finding all star (or unitary) representations
of the Lie superalgebra $\osp(1|2)$.  These are known, and are characterized
by a positive real number $\beta$ and a vacuum vector $|\beta,0\rangle$, such that
\begin{equation*}
b^-|\beta, 0\rangle = 0,\quad
\{b^-,b^+\} |\beta,0\rangle = 2\beta |\beta,0\rangle.
\end{equation*}
The representation space $\Gamma_\beta$ is the Hilbert space $\ell^2({\mathbb{Z}}_+)$
with orthonormal basis vectors $|\beta,n\rangle$ ($n\in {\mathbb{Z}}_+$) and with the following 
actions:
\begin{equation}
\begin{aligned}
	b^+ |\beta,2n\rangle & = \sqrt{2(n+\beta)}\,|\beta,2n+1\rangle, & & \quad &
	b^- |\beta,2n\rangle & = \sqrt{2n}\,|\beta,2n-1\rangle, \\ 
	b^+ |\beta,2n+1\rangle & = \sqrt{2(n+1)}\,|\beta,2n+2\rangle, & &  \quad & 
	b^- |\beta,2n+1\rangle & = \sqrt{2(n+\beta)}\,|\beta,2n\rangle,  
\end{aligned}
	\label{bpm-actions}
\end{equation}
from which follows
\begin{equation}
\{ b^-, b^+ \}|\beta,n\rangle = 2(n+\beta)\,|\beta,n\rangle,
	\label{action-anticomm}
\end{equation}
leading to the spectrum of $\hat H_0$. From the action of $\hat H_0$ and from the explicit action
of the commutator $[\hat q,\hat p]$ on the basis vectors $|\beta,n\rangle$, it is clear that for
$\beta=1/2$ the paraboson oscillator yields the canonical oscillator.
The position wavefunctions for the paraboson oscillator can then be determined by constructing the
formal eigenvectors of $\hat q$ (see~\cite{JSV2011}) or by different techniques~\cite{Mukunda}.

To see that the Lie superalgebra generated by the paraboson operators is indeed $\osp(1|2)$, let
\begin{equation}
H=\frac{1}{4} \{ b^-,b^+\}, \quad
E^+= \frac{1}{4} \{ b^+,b^+\}, \quad
E^-= -\frac{1}{4} \{ b^-,b^-\}; \quad
B^+= \frac{1}{2\sqrt{2}} b^+, \quad
B^-= \frac{1}{2\sqrt{2}} b^-.
\label{osp12basis}
\end{equation}
Then, using only \eqref{CCs2} one finds back the standard $\osp(1|2)$ commutation relations~\cite[p.~260]{Frappat}:
\begin{align}
&[H,E^\pm]=\pm E^\pm,\qquad [E^+,E^-]=2H,\qquad [H,B^\pm]=\pm\frac{1}{2} B^\pm, \nonumber\\
&[E^\pm,B^\mp]=-B^\pm, \qquad\{B^+,B^-\}=\frac{1}{2}H,\qquad  \{B^\pm, B^\pm\} = \pm\frac{1}{2} E^\pm.
\label{ospCR}
\end{align}
From these relations, the embedding of $\osp(1|2)$ into $\ssl(2|1)$ is also clear.
Starting from~\eqref{co-odd}-\eqref{co-mixed}, keeping $H$, $E^+$ and $E^-$ and putting
\begin{equation}
B^+=\frac{1}{2}(F^++G^+),\qquad B^-=\frac{1}{2}(F^- - G^-),
\label{BFG}
\end{equation}
one finds again the relations~\eqref{ospCR} (which is why we have used the same names for the corresponding generators).

It is now easy to verify that the $\ssl(2|1)$ irreducible representation $\Pi_\beta$ of section~2 decomposes, under
the embedding $\ssl(2|1)\supset \osp(1|2)$, as a single irreducible representation $\Gamma_\beta$ of $\osp(1|2)$.
Note that in this context the paraboson position operator $\hat q$ is given by
\begin{equation}
\hat q = \frac{1}{\sqrt{2}}(b^+ + b^-) = 2(B^+ + B^-) = F^++G^++F^--G^-.
\label{q-osp}
\end{equation}
Comparing with~\eqref{hatq} reveals why the case $\gamma=1$ corresponds to the paraboson oscillator, and why
the $\ssl(2|1)$ oscillator wavefunctions $\Phi_n^{(\beta,1)}(x)$ coincide with the paraboson oscillator wavefunctions.

\section{Remarks and discussion}

The oscillator models described by the discrete series representations $\Pi_\beta$ of the Lie superalgebra
$\ssl(2|1)$ offer an interesting extension of classical oscillator models.
They are characterized by two parameters: $\beta>0$ is a representation parameter, and $\gamma (\ne 0)$ is
an extra parameter appearing in the expression for the position operator $\hat q$.

Let us consider here some other quantities in these models that may play a role in physics.
First of all, using the expression of ${\hat q}$ and ${\hat p}$ and the actions~\eqref{FGact},
one finds in the representation $\Pi_\beta$:
\begin{align}
& [\hat q,\hat p ]\; |\beta,2n\rangle = 2\ii \left(\beta+(1-\gamma^2)n \right) |\beta,2n\rangle \qquad(n=0,1,2,\ldots), \nonumber\\
& [\hat q,\hat p ]\; |\beta,2n-1\rangle = 2\ii \left((1-\beta)-(1-\gamma^2)n \right) |\beta,2n-1\rangle \qquad(n=1,2,\ldots).
\label{pq}
\end{align}
So, just as for the paraboson oscillator, the action of the commutator $[\hat q,\hat p ]$ is still diagonal.
Note that for $\gamma^2=1$, one finds indeed
\begin{align*}
& [\hat q,\hat p ]\; |\beta,2n\rangle = 2\ii \beta |\beta,2n\rangle \qquad(n=0,1,2,\ldots), \nonumber\\
& [\hat q,\hat p ]\; |\beta,2n-1\rangle = 2\ii (1-\beta) |\beta,2n-1\rangle \qquad(n=1,2,\ldots),
\end{align*}
which is a known expression for the paraboson case. And clearly, for $\beta=1/2$, this 
becomes $[\hat q,\hat p ]=\ii$, the canonical situation.

Another operator that is worth considering in $\Pi_\beta$ is $\frac{{\hat p}^2}{2}+\frac{{\hat q}^2}{2}$,
since this stands for the Hamiltonian in the paraboson (and thus also in the canonical) case.
It is easy to verify that
\begin{align}
& (\frac{{\hat p}^2}{2}+\frac{{\hat q}^2}{2}) |\beta,n\rangle = 
\left( \frac{\gamma^2+1}{2} n +\beta \right)|\beta,n\rangle \qquad\hbox{for $n$ even}, \nonumber\\
& (\frac{{\hat p}^2}{2}+\frac{{\hat q}^2}{2}) |\beta,n\rangle = 
\left( \frac{\gamma^2+1}{2} n +\beta + \frac{\gamma^2-1}{2}\right)|\beta,n\rangle \qquad\hbox{for $n$ odd} .
\label{p2q2}
\end{align}
Again, this operator is diagonal in the general case, with a very simple action.
Clearly, for $\gamma^2=1$ this reduces to the paraboson oscillator Hamiltonian (or the Wigner oscillator).

To summarize, we have developed a new model for the quantum oscillator based upon the Lie superalgebra $\ssl(2|1)$
and its discrete series representations $\Pi_\beta$. 
These are infinite-dimensional unitary representations labeled by a positive number $\beta$, and the action of the
$\ssl(2|1)$ basis elements in this representation has been determined explicitly.
The Hamiltonian $\hat H$, the position $\hat q$ and the momentum $\hat p$ of the model are three self-adjoint elements
of $\ssl(2|1)$ satisfying the Hamilton-Lie equations~\eqref{Hqp}.
In particular, this requirement leaves a degree of freedom in the choice of the position operator, 
giving rise to an arbitrary parameter $\gamma$ in the expression for $\hat q$.
The spectrum of $\hat H$ coincides with that of the canonical quantum oscillator.
The spectrum of $\hat q$ depends on $\gamma$, and can be infinite discrete ($|\gamma|\ne 1$) or continuous ($|\gamma|=1$).
The position wavefunctions $\Phi_n^{(\beta,\gamma)}(x)$ have been determined explicitly.
For $\gamma=1$ they coincide with paraboson wavefunctions, and are given in terms of Laguerre polynomials.
In particular, the wavefunctions $\Phi_n^{(1/2,1)}(x)$ are those of the canonical quantum oscillator in terms
of Hermite polynomials.
For $|\gamma|\ne 1$ they are given in terms of Meixner polynomials, and satisfy a discrete orthogonality relation.
Plots of the discrete wavefunctions reveal properties that are very similar to those of the canonical oscillator (when
$\beta=1/2$) or to the paraboson oscillator (when $\beta>1/2$).
{}From the closely related momentum wavefunctions, the $\ssl(2|1)$ Fourier transform has been constructed
in explicit form.
The embedding of the Lie superalgebra $\osp(1|2)$ into $\ssl(2|1)$ offers an algebraic explanation of the appearance
of paraboson wavefunctions in the $\ssl(2|1)$ model, since $\osp(1|2)$ is the superalgebra underlying
the paraboson oscillator algebra. 

The wavefunctions appearing here are Meixner polynomials of the type $M_n(x^2)$ for even wavefunctions and
of the type $x M_n(x^2)$ for odd wavefunctions.
This kind of structure is reminiscent of some so-called ``$-1$ polynomials''.
The $-1$ polynomials are usually considered~\cite{Vinet2011} to be the $q=-1$ limit of basic hypergeometric polynomials 
appearing in the $q$-Askey scheme~\cite{Koekoek} (as long as this limit makes sense). 
As far as we know, they have not been explored systematically~\cite{Vinet2011}. 
The first example appeared in work of Bannai and Ito~\cite{Bannai}, where a $q=-1$ limit of $q$-Racah
polynomials plays a role. 
More recently, the $q=-1$ limit of little $q$-Jacobi polynomials was considered in~\cite{Vinet2011}, 
and the $q=-1$ limit of dual $q$-Hahn polynomials in~\cite{Tsujimoto2011}.
For these examples, expressing the $-1$ polynomials in terms of their classical counterparts $P_n(x)$, 
one sees indeed that they are essentially given in terms of $P_n(x^2)$ and $xP_n(x^2)$.

Note also that the underlying algebra $\ssl(2|1)$ has a natural coproduct, so it should be feasible to
construct tensor products of the representations given in this paper, and determine the corresponding Clebsch-Gordan
coefficients. 
Note that the representations $\Pi_\beta$ of $\ssl(2|1)$ coincide (see section~6) with the representations 
$\Gamma_\beta$ of $\osp(1|2)$, as far as the representation space and the action of $\osp(1|2)$ is concerned. 
These paraboson oscillator representations are essentially equal to the class of representations
of $\ssl_{-1}(2)$, considered in~\cite{Genest2012}. The Clebsch-Gordan problem for $\ssl_{-1}(2)$ has been solved
in~\cite{Genest2012}, and the Clebsch-Gordan coefficients are determined in terms of dual $-1$ Hahn polynomials.
It would be interesting to see whether the Clebsch-Gordan problem for $\ssl(2|1)$ has the same solution.
If that is the case, this might lead to interesting relations between dual $-1$ Hahn polynomials
and the Meixner polynomials appearing in this paper.

\section*{Acknowledgments}
E.I.~Jafarov was supported by a postdoc fellowship from the Azerbaijan National Academy of Sciences.

\newpage

\newpage

\begin{thebibliography}{99}

\bibitem{Andrews}
G.E.\ Andrews, R.\ Askey and R.\ Roy,
{\em Special functions}
(Cambridge University Press, Cambridge, 1999).

\bibitem{Arik1999}
M.\ Arik, N.M.\ Atakishiyev and K.B.\ Wolf,
%Quantum algebraic structures compatible with the harmonic oscillator Newton equation,
J.\ Phys.\ A {\bf 32}, L371–-L376 (1999).

\bibitem{AJNW1998}
N.M.\ Atakishiyev, E.I.\ Jafarov, S.M.\ Nagiyev and K.B.\ Wolf,
%Meixner oscillators,
Rev.\ Mex.\ Fis.\ {\bf 44}, 235-244 (1998).

%\bibitem{Atak1994}
%N.M.\ Atakishiyev and K.B.\ Wolf,
%Approximation on a finite set of points through Kravtchuk functions,
%Rev.\ Mex.\ Fis. {\bf  40}, 366-377 (1994).

%\bibitem{Atak1997}
%N.M.\ Atakishiyev and K.B.\ Wolf,
%Fractional Fourier-Kravtchuk transform,
%J.\ Opt.\ Soc.\ Am.\ A. {\bf  14}, 1467-1477 (1997).

%\bibitem{Atak1999b}
%N.M.\ Atakishiyev, L.E.\ Vicent and K.B.\ Wolf,
%Continuous versus Discrete Fractional Fourier Transforms,
%J.\ Comp.\ Appl.\ Math. {\bf  107}, 73-95 (1999).

\bibitem{Atak2001}
N.M.\ Atakishiyev, G.S.\ Pogosyan, L.E.\ Vicent and K.B.\ Wolf,
%Finite two-dimensional oscillator: I. The cartesian model,
J.\ Phys.\ A {\bf 34}, 9381-9398 (2001).

\bibitem{Atak2001b}
N.M.\ Atakishiyev, G.S.\ Pogosyan, L.E.\ Vicent and K.B.\ Wolf,
%Finite two-dimensional oscillator: I. The radial model,
J.\ Phys.\ A {\bf 34}, 9399-9415 (2001).

\bibitem{Atak2005}
N.M.\ Atakishiyev, G.S.\ Pogosyan and K.B.\ Wolf,
%Finite models of the oscillator,
Phys.\ Part.\ Nuclei {\bf  36}, 247-265 (2005).

\bibitem{Atak2006}
M.N.\ Atakishiyev, N.M.\ Atakishiyev and A.U.\ Klimyk,
%On $su_q(1,1)$-model of quantum oscillator,
J.\ Math.\ Phys. {\bf 47}, 093502 (2006).

\bibitem{Bailey}
W.N.\ Bailey,
{\em Generalized hypergeometric series} 
(Cambridge University Press, Cambridge, 1964).

\bibitem{Bannai}
E.\ Bannai and T.\ Ito,
{\em Algebraic Combinatorics I: Association Schemes}
(Benjamin \&\ Cummings, Mento Park CA, 1984).

\bibitem{Berezanskii}
Yu.\ M.\ Berezanski\u{\i},
{\em Expansions in eigenfunctions of selfadjoint operators}, 
{American Mathematical Society, Providence, 1968}.

\bibitem{Biedenharn1989}
L.C.\ Biedenharn,
%The quantum group ${\rm SU}_q(2)$ and a $q$-analogue of the boson operators.  
J.\ Phys.\ A {\bf 22}, L873-–L878 (1989).

\bibitem{Erdelyi}
A.\ Erd\'elyi, W.\ Magnus, F.\ Oberhettinger, F.G.\ Tricomi, 
{\em Higher Transcendental Functions}, Volume~1
(McGraw-Hill, New York, 1953).

\bibitem{Frappat}
L.\ Frappat, A.\ Sciarrino and P.\ Sorba,
{\em Dictionary on Lie Algebras and Superalgebras}
(Academic Press, London, 2000).

\bibitem{Ganchev}
A.C.\ Ganchev and T.D.\ Palev,
% A LIE SUPERALGEBRAIC INTERPRETATION OF THE PARA-BOSE STATISTICS,
J.\ Math.\ Phys. {\bf 21}, 797-799 (1980).

\bibitem{Genest2012}
V.X.\ Genest, L.\ Vinet and A.\ Zhedanov,
{\em The algebra of dual $-1$ Hahn polynomials and the Clebsch-Gordan problem of $sl_{-1}(2)$},
arXiv:1207.4220v1 [math-ph] (2012).

\bibitem{Green53}
H.S.\ Green,
%A Generalized Method of Field Quantization,
Phys.\ Rev. {\bf 90}, 270-273 (1953).

\bibitem{Ismail}
M.E.H.\ Ismail,
{\em Classical and quantum orthogonal polynomials in one variable}
(Cambridge University Press, Cambridge, 2005).

\bibitem{JLV2008}
E.\ Jafarov, S.\ Lievens and J.\ Van der Jeugt,
%The Wigner distribution function for the one-dimensional parabose oscillator.
J.\ Phys.\ A {\bf 41}, 235301 (2008).

\bibitem{JSV2011}
E.I.\ Jafarov, N.I.\ Stoilova and J.\ Van der Jeugt,
%Finite oscillator models: the Hahn oscillator,
J.\ Phys.\ A {\bf 44}, 265203 (2011).

\bibitem{JSV2011b}
E.I.\ Jafarov, N.I.\ Stoilova and J.\ Van der Jeugt,
%The su(2)a Hahn oscillator and a discrete Fourier-Hahn transform,
J.\ Phys.\ A {\bf 44}, 355205 (2011).

\bibitem{JSV2012}
E.I.\ Jafarov, N.I.\ Stoilova and J.\ Van der Jeugt,
%Deformed su(1,1) algebra as a model for quantum oscillators,
SIGMA {\bf 8}, 025 (2012).

\bibitem{JV2012}
E.I.\ Jafarov and J.\ Van der Jeugt,
%A finite oscillator model related to $\ssl(2|1)$,
J.\ Phys.\ A {\bf 45}, 275301 (2012).

\bibitem{King2003}
R.C.\ King, T.D.\ Palev, N.I.~Stoilova and J.\ Van der Jeugt,
% The non-commutative and discrete spatial structure of a 3D Wigner quantum oscillator
J.\ Phys.\ A: Math.\ Gen. {\bf 36}, 4337-4362 (2003).

\bibitem{Klimyk2005}
A.U.\ Klimyk,
%On position and momentum operators in the $q$-oscillator.  
J.\ Phys.\ A {\bf 38}, 4447-–4458 (2005).

\bibitem{Klimyk2006}
A.U.\ Klimyk,
% The su(1,1)-models of quantum oscillator,
Ukr.\ J.\ Phys. {\bf 51}(10), 1019-1027 (2006).

\bibitem{Koekoek}
R.\ Koekoek, P.A.\ Lesky and R.F.\ Swarttouw,
{\em Hypergeometric orthogonal polynomials and their $q$-analogues} 
(Springer-Verlag, Berlin, 2010).

\bibitem{Koelink2004}
H.T.\ Koelink, 
{\em Spectral theory and special functions}, 
in ``Laredo Lectures on Orthogonal Polynomials and Special Functions'', eds. R. \'Alvarez-Nodarse, F. Marcell\'an, W. Van Assche
(Nova Science Publishers, NY, 2004), p. 45-84.

\bibitem{Koelink1998}
H.T.\ Koelink and J.\ Van der Jeugt,
%Convolutions for orthogonal polynomials from Lie and quantum algebra representations,
SIAM J.\ Math.\ Anal. {\bf 29}, 794-822 (1998).

\bibitem{Macfarlane1989}
A.J.\ Macfarlane,
%On $q$-analogues of the quantum harmonic oscillator and the quantum group ${\rm SU}(2)_q$,
J.\ Phys.\ A {\bf 22}, 4581-4588 (1989).

\bibitem{Marcu}
M.\ Marcu,
%The representations of spl(2|1): an example of representations of basic superalgebras,
J.\ Math.\ Phys. {\bf 21}, 1277-1283 (1980).

\bibitem{Mukunda}
N.\ Mukunda, E.C.G.\ Sudarshan, J.K.\ Sharma and C.L.\ Mehta,
%REPRESENTATIONS AND PROPERTIES OF PARA-BOSE OSCILLATOR OPERATORS .1. ENERGY POSITION AND MOMENTUM EIGENSTATES,
J.\ Math.\ Phys. {\bf 21}, 2386-2394 (1980).

\bibitem{Ohnuki}
Y.\ Ohnuki and S.\ Kamefuchi,
{\em Quantum Field Theory and Parastatistics}
(Springer-Verlag, New-York, 1982).

\bibitem{Palev79}
T.D.\ Palev,
%  Lie-superalgebraical approach to the second quantization
Czech J.\ Phys., Sect. {\bf B29}, 91-98 (1979).

%\bibitem{Palev82}
%T.D.\ Palev,
% Wigner approach to quantization. Noncanonical quantization of two particles interacting via a harmonic potential 
%J.\ Math.\ Phys. {\bf 23}, 1778-1784 (1982).

\bibitem{Regniers2010}
G.\ Regniers and J.\ Van der Jeugt,
%Wigner quantization of some one-dimensional Hamiltonians. 
J.\ Math.\ Phys. {\bf 51}, 123515 (2010). 

\bibitem{Scheunert1977}
M.\ Scheunert, W.\ Nahm and V.\ Rittenberg,
% Irreducible representations of the osp(1,2) and spl(1,2) graded Lie algebras,
J.\ Math.\ Phys. {\bf 18}, 155-162 (1977).

\bibitem{Slater}
L.J.\ Slater,
{\em Generalized hypergeometric functions} 
(Cambridge University Press, Cambridge, 1966).

\bibitem{Sun1989}
Chang-Pu Sun and Hong-Chen Fu, 
%The q-deformed boson realisation of the quantum group $SU(n)_q$ and its representations, 
J.\ Phys.\ A {\bf 22}, L983-L988 (1989). 

\bibitem{Tsujimoto2011}
S.\ Tsujimoto, L.\ Vinet and A.\ Zhedanov,
{\em Dual $-1$ Hahn polynomials: ``classical'' polynomials beyond the Leonard duality,}
arXiv: 1108.0132 [math.CA] (2011).

\bibitem{Jagan1998}
J.\ Van der Jeugt and R.\ Jagannathan,
%Realizations of su(1,1) and U_q(su(1,1)) and generating functions for orthogonal polynomials,
J.\ Math.\ Phys. {\bf 39}, 5062-5078 (1998). 

\bibitem{Vinet2011}
L.\ Vinet and A.\ Zhedanov,
%{\em A ``missing'' family of classical orthogonal polynomials},
J.\ Phys.\ A {\bf 44}, 085201 (2011).

\bibitem{Wigner}
E. P.\ Wigner, 
% Do the equations of motion determine the quantum mechanical commutation relations? 
Phys.\ Rev. {\bf 77}, 711-712 (1950).


\end{thebibliography}
\end{document}